\documentclass[aps,twocolumn,
showpacs,pra,amssymb, amsmath]{revtex4}
\usepackage{amsmath,latexsym,amssymb}
\usepackage[dvips]{graphicx}
\usepackage{amsmath}
\usepackage{latexsym}
\usepackage{amssymb}
\usepackage{bm}

\begin{document}

\title{Comment on ``All quantum observables in a hidden-variable model must
commute simultaneously"}
\author{Koji Nagata}
\affiliation{National Institute of Information and Communications 
Technology, 4-2-1 Nukuikita, Koganei, Tokyo 184-8795, Japan}
\date{\today}

\begin{abstract}
Malley discussed {[Phys. Rev. A {\bf 69}, 022118 (2004)]} 
that all quantum observables 
in a hidden-variable model for quantum events must commute simultaneously.
In this comment, we discuss that 
Malley's theorem is indeed
valid for the hidden-variable theoretical assumptions, which were 
introduced by Kochen and Specker.
However, we give an example that the local hidden-variable (LHV)
model for quantum events
preserves noncommutativity of quantum observables.
It turns out that Malley's theorem is not 
related to the LHV model for quantum events, in general. 
\end{abstract}

\maketitle

\section{Introduction}

Physical observables do not generally commute 
in the Hilbert space formalism of quantum theory \cite{bib:Redhead,bib:Peres}.
Recently, Malley 
discussed \cite{bib:Malley1,bib:Malley2} that all 
quantum observables must 
commute simultaneously if 
we accept a hidden-variable (HV) model for quantum events.

First, Malley showed that all 
quantum observables must 
commute simultaneously under a special set 
of assumptions valid for some HV model for quantum events.
According to Malley's paper, 
the special set 
of assumptions is equivalent to 
those under which the Kochen-Specker (KS) theorem \cite{bib:KS} is derived.
And, Malley claimed that these conditions are 
also equivalent to those under which the 
Bell inequalities \cite{bib:Bell} are derived, upon invoking Fine's paper 
--proposition (2) in Ref.~\cite{bib:Fine}.
Finally, Malley concluded that the experimental 
violations of the Bell inequalities demonstrate 
only that quantum observables do not commute.

One can find that the argument by 
Malley is indeed valid under the special HV 
theoretical assumptions which were used 
in order to construct Malley's theorem.
In more detail, 
the product rule (the KS condition) and 
the uniqueness feature of Gleason's theorem 
imply that all quantum observables commute simultaneously \cite{bib:Malley2}.

On the other hand, another type of model 
for quantum events has been 
presented \cite{bib:Khrennikov1,bib:Khrennikov2}.
It can be interpreted by a HV model. 
The HV model for 
quantum events says that classical random variables are 
represented by in general noncommutative operators 
in the Hilbert space formalism of quantum theory.
In other words, the HV model for quantum 
events preserves noncommutativity of 
quantum observables, even though we accept it.

By reading Bell's arguments \cite{bib:Bell} carefully, we can 
notice the condition under which Bell's theorem is derived.
In fact, the condition is only that 
quantum correlation functions are reproducible by the 
classical-random-variables model for quantum events.
Further, a classical random variable related to one site 
must not depend on the (simultaneous) choices of 
measurement observables on
the other site each other.
(This condition is related to Bell's nonlocality.)
Namely, we can say that 
the local hidden-variable (LHV) model for quantum events 
was constructed only by classical random variables. And they depend on 
quantum measurement observables (Hermitian operators) for each sites.
Hence, one 
can see that the LHV model for quantum events can coexist with the HV model 
for quantum events
presented in Refs.~\cite{bib:Khrennikov1,bib:Khrennikov2}.

Therefore, we may be confused.
Do all 
quantum observables 
commute simultaneously when 
we accept the LHV  model for quantum events?
Do the experimental 
violations of the Bell inequalities demonstrate 
only that quantum observables do not commute?
In fact, this problem was also discussed in Ref.~\cite{bib:Fine2}
from different approach (Theorem 7 in Ref.~\cite{bib:Fine2}).

We shall investigate the reason why the argument claimed by Malley 
gives rise to 
the contradiction against the existence of the HV model presented in 
Refs.~\cite{bib:Khrennikov1,bib:Khrennikov2}.
In what follows, we shall give an example such that the LHV
model for quantum events preserves 
noncommutativity of quantum observables, on using several quantum states.

\section{LHV model suggesting noncommutativity }

In what follows, we shall mention the standard 
LHV model for quantum events.
And we shall give a counterexample against Malley's claim.

Let $L(H)$
be the space of Hermitian operators acting on a finite-dimensional Hilbert
space $H$, and $T(H)$ 
be the space of density operators
acting on the Hilbert
space $H$. Namely, $T(H)=
\{\rho | \rho\in L(H)\wedge\rho\geq 0\wedge {\rm tr}[\rho]=1\}$.

Let us 
consider a classical probability space $(\Omega,\Sigma,M_{\rho})$, where
$\Omega$ is a nonempty space, $\Sigma$ is a
$\sigma$-algebra of subsets of $\Omega$, and $M_{\rho}$ is a
$\sigma$-additive normalized measure on $\Sigma$ such that 
$M_{\rho}(\Omega)=1$.
The subscript $\rho$ expresses the following meaning:
The probability measure $M_{\rho}$ is determined uniquely
when the state $\rho$ is specified.

Consider bipartite states $\rho$ 
in $T(H_1\otimes H_2)$,
where $H_k$ represents the Hilbert space with respect to 
party $k=1,2$.

Then we can
define functions
$f_k: v_k,\omega \mapsto 
f_k(v_k,\omega)\in 
{[I(v_k),S(v_k)]},
v_k\in{L}({H}_k), 
\omega\in \Omega$.
Here, $S(v_k)$ and $I(v_k)$ are the supremum and the infimum of the spectrum 
of Hermitian operators $v_k$, respectively.

The
functions $f_k(v_k,\omega)$ 
must not depend on the choices of $v$'s on
the other site each other.
On using the functions $f_k$, 
we can define quantum states which admit the LHV model \cite{bib:Werner1}.
Namely, a quantum state is said to admit the LHV model
if and only if 
there exist a classical probability space $(\Omega,\Sigma,M_{\rho})$ and
a set of functions $f_1,f_2$,  
such that
\begin{eqnarray}
\int_{\Omega}\!\! M_{\rho}(d\omega)
f_1(v_1,\omega)f_2(v_2,\omega)
={\rm tr}[\rho v_1\otimes v_2],\label{LHV}
\end{eqnarray}
for every
Hermitian operator in the following form: $v_1\otimes v_2$.
Here, $v_k
\in L(H_k)$.
Note that there are several
(noncommuting) observables per site (not just one $v_k$).

The meaning of Eq.~(\ref{LHV}) is as follows:
All correlation functions ${\rm tr}[\rho v_1\otimes v_2]$ 
in the state $\rho$
are reproducible by the LHV model for quantum events.

Let us consider the Pauli spin-1/2 
operators, $\sigma^k_x,\sigma^k_y,$ and $\sigma^k_z$.
Let us assume the system is in an element of certain set of 
two-spin-1/2 states. 
They are bipartite 
separable states written by
\begin{eqnarray}
U(\alpha, \beta)=\alpha|+_1,+_2\rangle\langle+_1,+_2|
+\beta|-_1,-_2\rangle\langle-_1,-_2|\label{uncorre}
\end{eqnarray}
where $\sigma^k_z|\pm_k\rangle=\pm 1|\pm_k\rangle$ and 
$\alpha+\beta=1, \alpha, \beta\geq 0$.

As is well known, every separable state admits 
the LHV model \cite{bib:Werner1}.
In other words, all correlation functions in those 
separable states $U(\alpha, \beta)$ 
are described
with the property that they are reproducible by 
the LHV model for quantum events.
Hence functions ($f_1, f_2$) exist.
That is, we obtain the following equation:
\begin{eqnarray}
\int_{\Omega}\!\! M_{U(\alpha, \beta)}(d\omega)
f_1(v_1,\omega)f_2(v_2,\omega)
={\rm tr}[U(\alpha, \beta) v_1\otimes v_2]\label{L}
\end{eqnarray}
for every observable $v_1\otimes v_2$ and every $\alpha, \beta$.
From Eq.~(\ref{L}), when $v_k=i[\sigma^k_x,\sigma^k_y]$, we have
\begin{eqnarray}
&&\int_{\Omega}\!\! M_{U(\alpha, \beta)}(d\omega)
f_1(i[\sigma^1_x,\sigma^1_y],\omega)
f_2(i[\sigma^2_x,\sigma^2_y],\omega)\nonumber\\
&&={\rm tr}[U(\alpha, \beta) i[\sigma^1_x,\sigma^1_y]\otimes 
i[\sigma^2_x,\sigma^2_y]].\label{LHVU}
\end{eqnarray}
Please notice that $i[\sigma^k_x,\sigma^k_y]=-2\sigma_z^k$ are
Hermitian operators.
On substituting Eq.~(\ref{uncorre}) into 
Eq.~(\ref{LHVU}) and performing some algebra, we find that 
\begin{eqnarray}
\int_{\Omega}\!\! M_{U(\alpha,\beta)}(d\omega)
f_1(i[\sigma^1_x,\sigma^1_y],\omega)
f_2(i[\sigma^2_x,\sigma^2_y],\omega)
=4
(\neq 0)\label{concl}
\end{eqnarray}
in spite of any possible values of $\alpha$ and of $\beta$.
This implies that there exists an 
event $(\Sigma', M_{U(\alpha, \beta)}(\Sigma')\neq 0)$ for 
a $\sigma$-algebra $\Sigma$ such that
\begin{eqnarray}
[\sigma^1_x, \sigma^1_y]\neq {\bf 0}
\wedge[\sigma^2_x, \sigma^2_y]\neq {\bf 0}
: \forall \omega \in \Sigma'
\end{eqnarray}
since 
$f_k({\bf 0},\omega)=0$
holds.
Here, ${\bf 0}$ represents the null operator.
We have assumed that the system is in an element of the set of
the states $U(\alpha, \beta)$.
But, the conclusion 
is independent of the possible values of $\alpha$ and of $\beta$.
Hence, there exist several quantum events for which the LHV model 
preserves noncommutativity of quantum observables.
Of course, no element of the set of the states $U(\alpha, \beta)$ 
says any violation of the Bell inequalities.
This fact gives rise to the conflict against
Malley's claim.

The experimental 
violations of the Bell inequalities demonstrate indeed
that quantum observables do not commute (Theorem 7 in Ref.~\cite{bib:Fine2}).
But, such violations show also the nonexistence
of the classical-random-variables model (i.e, 
the LHV model) for quantum events.
Therefore, one can see that such violations show also the nonexistence of the HV model proposed in Refs.~\cite{bib:Khrennikov1,bib:Khrennikov2}
(except for any nonlocal HV model even 
if the model proposed in Refs.~\cite{bib:Khrennikov1,bib:Khrennikov2}
could be applicable not only to the LHV model
but also to some nonlocal HV model.)


\section {Summary and discussion}

We have pointed out a contradiction.
That is, the argumentation presented in 
Refs.~\cite{bib:Khrennikov1,bib:Khrennikov2} 
cannot coexist with the argumentation claimed by Malley.
And we have given a counterexample against Malley's claim.
Namely, the LHV model for quantum events exists.
And it preserves noncommutativity of quantum observables.

From these arguments mentioned above, one can see that  
Malley's theorem is indeed true under special assumptions. 
In more detail, 
the product rule (the KS condition) and 
the uniqueness feature of Gleason's theorem 
imply that all quantum observables commute simultaneously \cite{bib:Malley2}.
In this sense, Malley's theorem is 
valid for the KS type of HV model 
for quantum events.
It was introduced by Kochen and Specker.
However, Malley's theorem is not 
related to the LHV model for quantum events, in general.
It was reported by Bell in 1964.

At the end of Malley's paper, 
it was discussed about {\it hybrid HV models}.
And Malley stated that ``violations 
of the Bell inequalities do not constitute a failure of Bell 
locality and 
our no-go commutativity 
result does not extend to a negation of Bell locality''.
However, the author thinks that the experimental violations of the Bell inequalities indeed constitute the failure of Bell's locality discussed in 1964.
Malley has explicitly written that the experimental violations of the Bell inequalities demonstrate only that quantum observables do not commute.
Hence, it seems that Malley considered that the no-go commutativity result can extend to the negation of original Bell's locality reported in 1964.
But this is not true,
because, there exists the explicit LHV model which is
compatible with noncommutative observables as we have shown.

It would be worth mentioning that 
the conclusion discussed in this comment coexists with 
the explicit difference between 
the KS theorem and Bell's theorem
in the Hilbert space formalism of quantum theory 
\cite{bib:Nagata1,bib:Nagata2}.
This approach is valid only when we are assumed to be given 
an arbitrary single state.
But, this kind of approach can be seen often in literature
\cite{bib:Redhead,bib:Peres,bib:Malley1,bib:Malley2,bib:KS,bib:Bell,bib:Khrennikov1,bib:Khrennikov2,bib:Fine,bib:Fine2,bib:Werner1,bib:Malley3,bib:Neumann}.

The author suspects that what Theorem 7 in Ref.~\cite{bib:Fine2} said is
as follows: 
Quantum observables must commute simultaneously if and only if 
we introduce a condition. 
The condition is that a set of quantum observables
 in the Hilbert space formalism
is isomorphic to a set of 
classical random variables which are defined on a common space.
Such a set of classical random variables 
obeys the classical (commutative) algebraic structure.
A similar approach has been seen in von Neumann's no-hidden-variables 
theorem \cite{bib:Neumann}.
Please notice that the outcome of 
the set of von Neumann's assumptions directly tells
that the set of all quantum observables 
cannot be isomorphic to any set obeying the classical algebraic structure.
In this mathematical sense, the fact 
that quantum observables, in general, do not commute 
(i.e., cannot be isomorphic to any set obeying the classical algebraic structure) is 
equivalent to 
von Neumann's no-hidden-variables theorem.
This fact agrees with Malley's achievements.
Finally, we mention that any nonlocal HV model 
was not taken into account.

\acknowledgments

The author would like to thank Professor Christoph Bruder for
valuable comments.



\end{document}